# An exact inviscid drag-adjoint solution for subcritical flows

Carlos Lozano[*], Jorge Ponsin[†]

National Institute of Aerospace Technology (INTA), Torrejón de Ardoz, Spain

## Introduction

This paper describes an exact solution to the drag-based adjoint Euler equations in two and three dimensions that is valid for subcritical, non-lifting irrotational flows and approximate for lifting flows.

The last thirty years have seen considerable progress in the use of adjoint equations in CFD for shape design [1], error estimation and mesh adaptation [2], stability analysis [3], etc. From the computational viewpoint, adjoint methods can be devised in two ways [4], which differ in how the discretized adjoint equations are obtained. In the continuous approach, one discretizes the adjoint pde, while in the discrete approach the adjoint equations are obtained directly from the discretized flow equations.

An important step in the application of adjoint methods is the development and verification of adjoint codes. This task is complicated by the lack of benchmark test cases, including exact solutions, and verification is usually done indirectly by comparing sensitivities with finite differences, which may lead to erroneous conclusions as rather accurate gradients can be derived from manifestly inaccurate adjoint solutions [5], or (in the case of discrete adjoint solvers) by cross-checking with the linearized solver [6].

A few exact solutions to the adjoint equations are known. For quasi-1D inviscid flows, the Green's function approach has provided a tool to generate exact adjoint solutions [7] [8] in terms of the base flow solution. For inviscid 2D/3D flows, the entropy variables $\partial_U S$ (the gradient of the entropy function with respect to the conservative flow variables), provide another exact solution for an output measuring net entropy flux across boundaries (including shock loci) [9] [10].

In this paper, we present another closed-form adjoint solution that corresponds to an output that measures aerodynamic drag and that is exact for 2D and 3D inviscid, irrotational, non-lifting flows, and approximate for lifting flows, and which turns out to be closely related to the entropy adjoint. A separate solution for inviscid, incompressible, irrotational flow is also presented, which is also exact for non-lifting flow. These solutions provide a cheap adjoint field for design or adaptation purposes that, perhaps more importantly in practical terms, can also serve to verify and debug adjoint solvers.

---

[*] Research Scientist, Computational Aerodynamics Group. Corresponding author: lozanorc@inta.es
[†] Research Scientist, Computational Aerodynamics Group.

## The drag-adjoint Euler equations

We begin by recalling a few facts regarding the inviscid (compressible) adjoint equations. We will focus, for definiteness, on steady, two-dimensional, inviscid flow on a domain $D$ with far-field boundary $S_\infty$ and wall boundary $S$ (typically an airfoil profile). Generalization to 3D is immediate. The flow is governed by the Euler equations $R(U) = \nabla \cdot \vec{F}(U) = 0$, where $U = (\rho, \rho\vec{v}, \rho E)^T$ are the conservative flow variables, $\vec{F} = (\rho\vec{v}, \rho\vec{v}u + p\hat{x}, \rho\vec{v}v + p\hat{y}, \rho\vec{v}H)^T$ is the flux vector and $\rho$, $\vec{v} = (u,v)$, $p$, $E$ and $H$ are the fluid's density, velocity, pressure, total energy and total enthalpy $H$, respectively. The adjoint equations are defined with respect to a functional of the flow variables, or cost function, that we take to be the drag coefficient

$$I = \int_S C_p (\hat{n} \cdot \vec{d}) ds \tag{1}$$

where $\vec{d} = (\cos\alpha, \sin\alpha)$, with $\alpha$ being the angle of attack, $\hat{n}$ is the outward unit normal vector at the boundary, $C_p = (p - p_\infty)/c_\infty$ is the non-dimensional pressure coefficient and $c_\infty$ is a normalization factor.

The sensitivities of the cost function (1) with respect to deformations $\delta\vec{x}$ of $S$ can be efficiently obtained with the continuous adjoint approach [1] (see [11] [12] [13] for more details). For subcritical flow, the adjoint variables $\psi = (\psi_1, \psi_2, \psi_3, \psi_4)^T$ obey the following adjoint equations and boundary conditions

$$\begin{aligned}\partial_U \vec{F}^T \cdot \nabla \psi &= 0 & \text{in } D \\ \hat{n} \cdot \vec{\varphi} &= c_\infty^{-1}(\vec{d} \cdot \hat{n}) & \text{at } S \\ \psi^T \hat{n} \cdot \partial_U \vec{F} \delta U &= 0 & \text{at } S_\infty\end{aligned} \tag{2}$$

where $\vec{\varphi} = (\psi_2, \psi_3)$. The same equations apply for incompressible flow (only that now the adjoint state contains 3 variables in 2D) with the only replacement of the Jacobian

$$\partial_U \vec{F} = \partial(\rho\vec{v}, \rho\vec{v}u + p\hat{x}, \rho\vec{v}v + p\hat{y})^T / \partial(p, \rho u, \rho v)$$

where $\rho$ is now constant.

# Analytic solution for the drag adjoint equations

## Compressible flow

Solving Eq. (2) in full generality is a formidable task that is only amenable to numerical computation. However, we have noticed that plots of drag adjoint solutions for subcritical cases show a striking resemblance to flow variables (see Fig. 1)

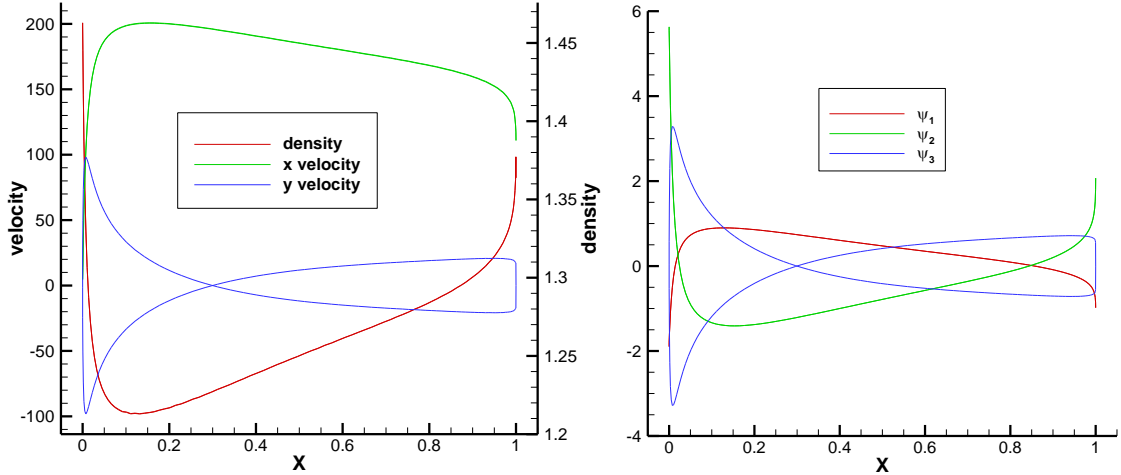

**Fig. 1 NACA0012, M = 0.5, α = 0°. Flow (left) and adjoint (right) variables computed with DLR's Tau code [14].**

We see that $\psi_1, \psi_2, \psi_3$ look like inverted, shifted and rescaled copies of $\rho, u, v$, respectively, and this resemblance extends throughout the flow domain. (For this cost function, $\psi_4 = \psi_1 / H$ is just a rescaled copy of $\psi_1$ [15], so we won't discuss it here). With that in mind, it is a matter of trial and error to guess an expression for the adjoint solution in terms of flow variables that obeys the adjoint equations and boundary conditions. It turns out that the choice

$$\psi^* = \frac{1}{\rho_\infty^{1-\gamma} q_\infty c_\infty} \begin{pmatrix} H\rho^{1-\gamma} - H_\infty \rho_\infty^{1-\gamma} \\ -\rho^{1-\gamma} u + \rho_\infty^{1-\gamma} q_\infty \cos\alpha \\ -\rho^{1-\gamma} v + \rho_\infty^{1-\gamma} q_\infty \sin\alpha \\ \rho^{1-\gamma} - \rho_\infty^{1-\gamma} \end{pmatrix} \qquad (3)$$

where $q_\infty = |\vec{v}_\infty|$, obeys the adjoint equations $\partial_U \vec{F}^T \cdot \nabla \psi^* = 0$ for homentropic flows, *i.e.*, flows that obey $\nabla(p/\rho^\gamma) = 0$. (3) also obeys the adjoint wall boundary condition $\hat{n} \cdot \vec{\varphi}^* = (\vec{d} \cdot \hat{n})/c_\infty$, as well as the far-field boundary conditions $(\psi^*)^T \hat{n} \cdot \partial_U \vec{F} \delta U = 0$ provided that the total enthalpy is constant and that the flow approaches a uniform (constant) far-field state $(\rho_\infty, q_\infty \cos\alpha, q_\infty \sin\alpha, H_\infty)$. This assumption does not hold for lifting flows [16] (though the solution could still be approximately valid in those situations), and prevents its application to more involved set-ups (such as the Ringleb flow which, being one of the few non-trivial problems of the 2D Euler equations for which an analytical solution is known, would make an ideal candidate here).

Hence, for non-lifting, homentropic and homenthalpic (and, thus, irrotational) inviscid flows, (3) is the exact solution for the drag-based adjoint equations. We will confirm this with several numerical tests in the next section, and investigate its applicability to lifting flows as well. The above results carry over to 3D with obvious modifications, although care must be exercised since in 3D, unlike the 2D case, flow past a lifting wing is necessarily rotational, and even a non-lifting wing may shed net vorticity in some spanwise regions, which must nevertheless cancel globally on one wing half.

Before moving on to the incompressible case, we note that eq. (3) corresponds to a near-field computation of drag (1). The corresponding exact solution for the far-field drag adjoint can be obtained as $\psi_{far} = \psi_{near} - (0, c_{\infty}^{-1} \cos\alpha, c_{\infty}^{-1} \sin\alpha, 0)^T$ [17].

Finally, it has been pointed out to us by one of the anonymous reviewers that the analytic adjoint solution (3) is actually equivalent to the entropy adjoint of [18]. Indeed, choosing the entropy function $S = -\rho \log(p/\rho^{\gamma})/(\gamma-1)$, where $\gamma$ is the adiabatic exponent, and taking into account that for isentropic flow $p_{\infty}/\rho_{\infty}^{\gamma} = p/\rho^{\gamma}$, it turns out that the entropy variables

$$\partial_U S = \left( \frac{\gamma - \log(p/\rho^{\gamma})}{\gamma - 1} - \frac{\rho \vec{v}^2}{2p}, \frac{\rho u}{p}, \frac{\rho v}{p}, -\frac{\rho}{p} \right)$$

are identical (up to a constant shift and rescaling) to (3)

$$\psi^* = -\frac{p_{\infty}}{\rho_{\infty} q_{\infty} c_{\infty}} (\partial_U S - \partial_U S_{\infty})^T$$

This relation can be understood if we recall that entropy variables are also identical, for subcritical flows, to the Oswatitsch adjoint [10], which is dual to drag measured as integrated entropy flux in the domain outer boundaries. This relation of drag to entropy variables had been pointed out and exploited before in [17].

Incompressible flow

For incompressible flow, a similar analysis as above yields the adjoint vector

$$\psi^* = \frac{1}{q_\infty c_\infty} \begin{pmatrix} 2(p_\infty - p)/\rho \\ q_\infty \cos\alpha - u \\ q_\infty \sin\alpha - v \end{pmatrix} \quad (4)$$

which is the analytic drag adjoint solution for irrotational base flows that approach the constant state $(p_\infty, q_\infty \cos\alpha, q_\infty \sin\alpha)$ at the outer boundary. Extension to 3D is again straightforward.

## Numerical tests

In this section, we are interested in verifying the validity of the proposed solution (3) on several test cases, including two and three-dimensional subsonic, non-lifting cases (for which (3) is an exact solution), subsonic, lifting cases (for which (3) may hold approximately), as well as a 2D transonic case (for which (3) is definitely not a solution) and an incompressible 2D case in order to test (4).

### Compressible flow

We first test a non-lifting case, namely steady inviscid flow past a NACA0012 airfoil at freestream Mach number $M_\infty = 0.5$ and angle of attack $\alpha = 0$. The numerical computations have been carried out with DLR's Tau solver with a central scheme with JST-type artificial dissipation on an unstructured triangular mesh with about 5200 nodes overall, and 200 nodes on the airfoil profile. Fig. 2 compares the numerical adjoint solution obtained with Tau's discrete (left) and continuous (right) adjoint solver against Eq. (3) evaluated with the corresponding numerical flow solution. It can be seen that both are in nearly perfect agreement with each other. We should however keep in mind that (3) is evaluated with a numerical flow solution, which may differ from the analytic solution (particularly at the trailing edge zone, where the flow solution should present a stagnation point), so the numerical adjoint solution is not necessarily following the analytic adjoint solution either.

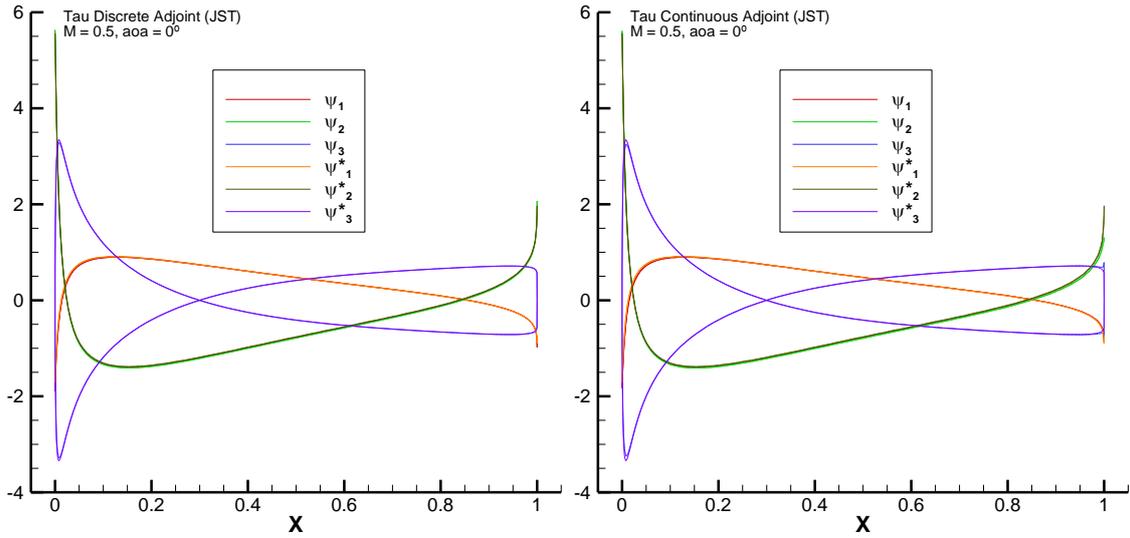

**Fig. 2** NACA0012, M = 0.5, α = 0º. Numerical ($\psi$) vs. analytic ($\psi^*$) drag adjoint variables computed with DLR's Tau continuous and discrete adjoint solvers.

As stated above, we do not expect (3) to hold exactly for a lifting case, so we repeat the above test but with α = 2º, leaving all other parameters (including the mesh) unchanged. As can be seen in Fig. 3, the ansatz (3) is quite approximately valid in this case, even though there are some clear discrepancies for the *x*-adjoint velocity ($\psi_2$). Thus, the benchmark solution can still be used to make a rough verification of the adjoint solution.

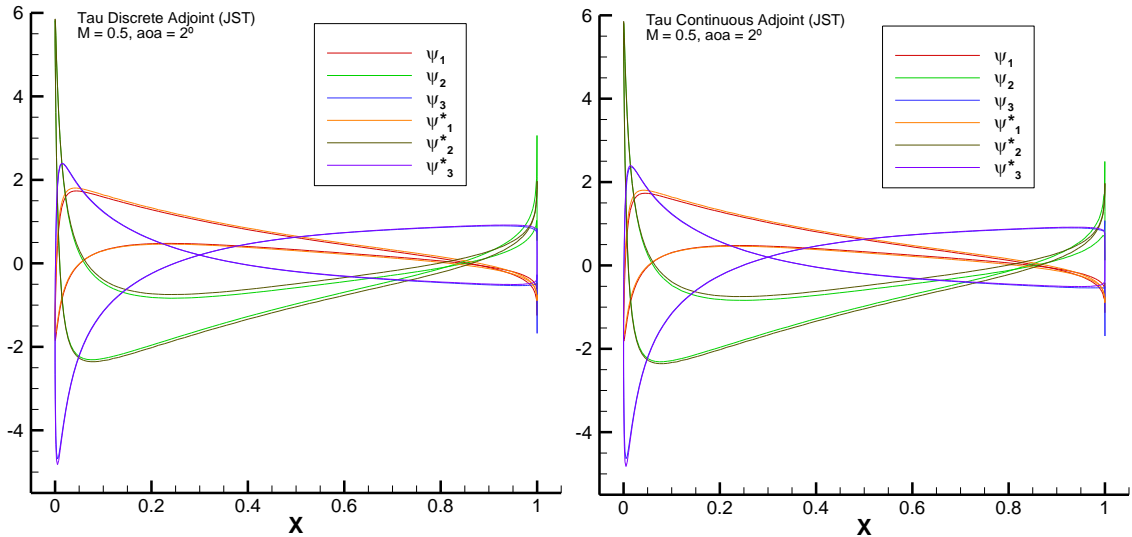

**Fig. 3** NACA0012, M = 0.5, α = 2º. Numerical ($\psi$) vs. analytic ($\psi^*$) drag adjoint variables.

Next, we examine a transonic, non-lifting case (NACA0012 with $M_\infty = 0.8$ and α = 0). Here the ansatz (3) is not expected to hold at all, and this is confirmed by Fig. 4. While the *y*-adjoint velocity ($\psi_3$) does show an overall similar underlying tendency as the flow solution, it is clear that flow solution cannot reproduce the singularities of the adjoint solution at the sonic points and, conversely, the adjoint solution cannot reproduce the strong shocks of the flow solution.

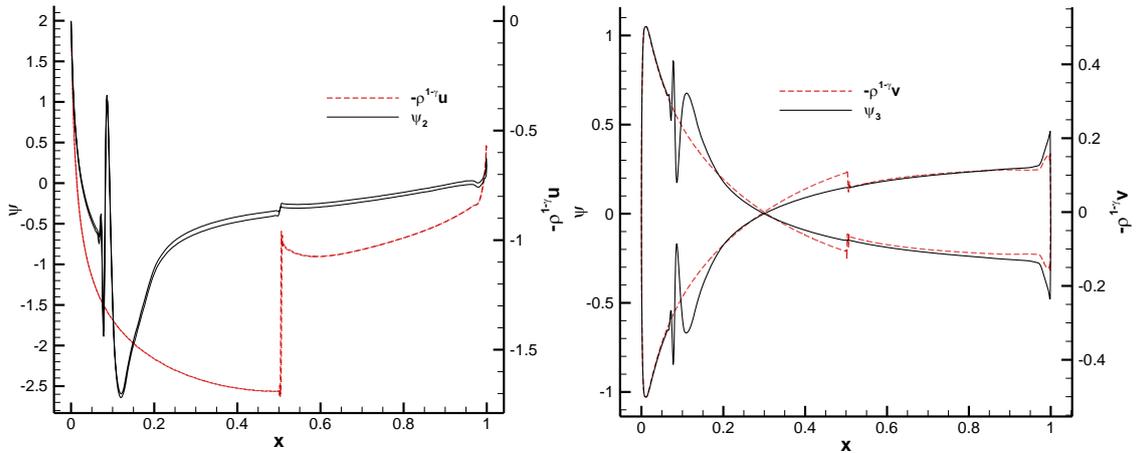

**Fig. 4 NACA0012, M = 0.8, α = 0°. Comparison of drag adjoint ($\psi$) and flow variables.**

In order to test Eq. (3) on a 3D flow, we consider flow over an ONERA M6 wing with $M_\infty = 0.54$, $\alpha = 0°$ and sideslip angle $\beta = 0°$, which results in fully subsonic flow. Likewise, since the wing sections are symmetric with respect to the chord and the wing has no twist, then for a subcritical, inviscid flow there should be no lift and no drag as the pressure forces acting on the wing cancel and, thus, no shed vorticity. Computations are carried out with Tau's discrete adjoint code on an unstructured mesh with 1.6 million tetrahedral cells with a central JST scheme. Fig. 5 compares the numerical adjoint solution with Eq. (3) at a spanwise section at 40% span, showing again a quite good agreement in all adjoint variables (we show only $\psi_1$, which corresponds to the density equation, and $\psi_4$, which corresponds to the $z$ (wall normal) momentum equations.

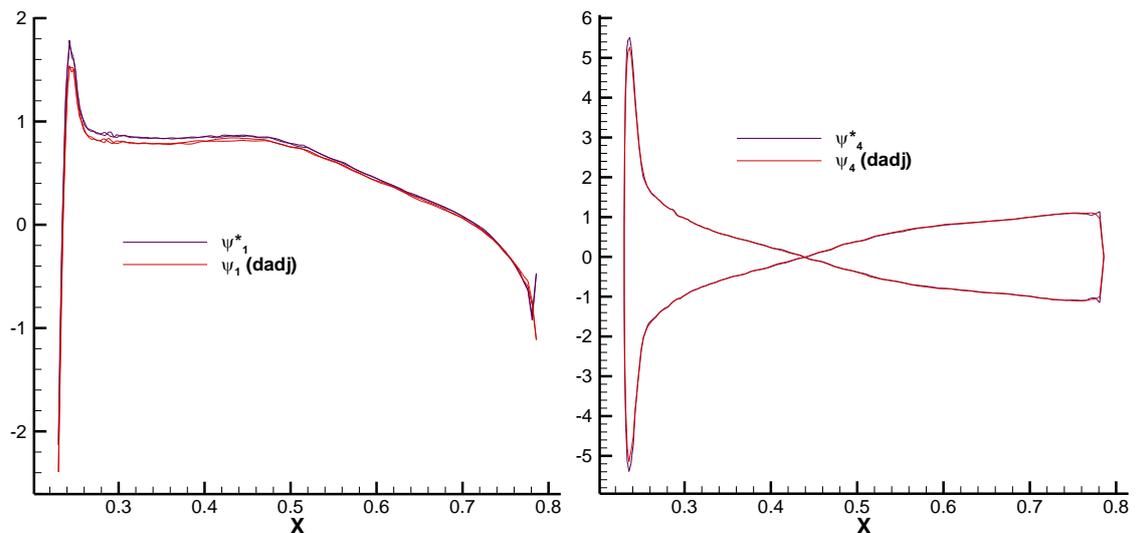

**Fig. 5 ONERAM6 wing (M = 0.54, α = 0°, β = 0°). Numerical ($\psi$) vs. analytic ($\psi^*$) drag adjoint variables on a spanwise section at 40% span.**

On the other hand, Fig. 6 shows that in a lifting case (the same ONERAM6 wing with $M_\infty = 0.54$, $\alpha = 3.06°$ and $\beta = 0°$), which is now necessarily rotational, there are clear discrepancies even though (as in 2D) the agreement in the wall-normal adjoint variable is remarkable.

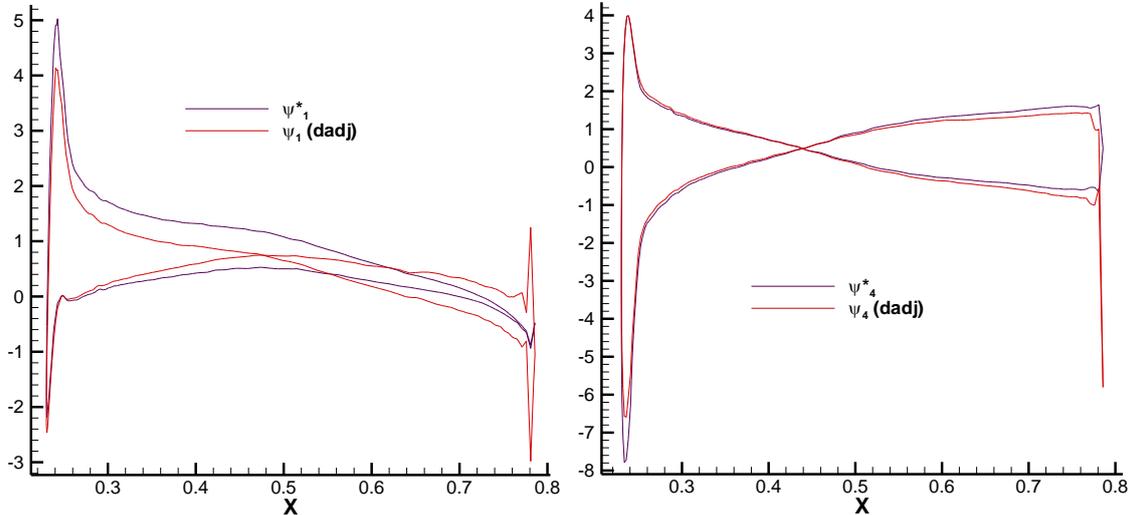

**Fig. 6 ONERAM6 wing (M = 0.54, α = 3.06º, β = 0º). Numerical ($\psi$) vs. analytic ($\psi^*$) drag adjoint variables on a spanwise section at 40% span.**

Incompressible flow

As our final test, we consider inviscid incompressible flow past a NACA0012 airfoil at zero angle of attack. The numerical computations have been carried out on the same mesh as above with SU2's incompressible continuous adjoint solver with a Roe-type upwind scheme [19]. The agreement between the numerical and benchmark adjoint solutions is again excellent.

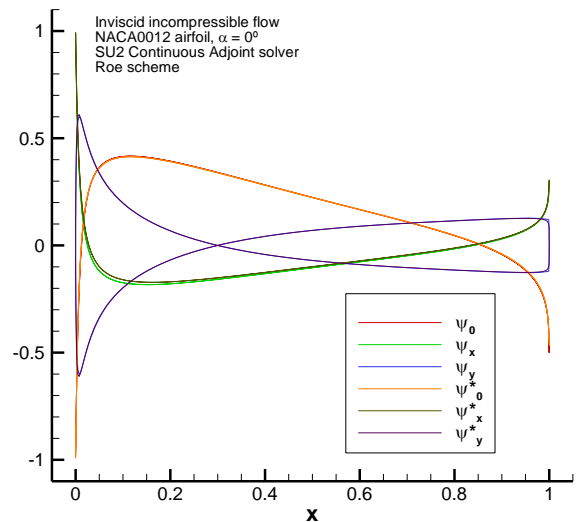

**Fig. 7 Numerical ($\psi$) vs. analytic ($\psi^*$) drag adjoint variables for incompressible inviscid flow past a NACA0012 with α = 0º.**

## Conclusions

We have described closed-form solutions for the drag-based adjoint equations for inviscid, irrotational, non-lifting flows. Even though no exact flow solution for such flows is generally available and, thus, the benchmark adjoint solution has to be constructed from numerical flow fields, it turns out that the agreement is in general excellent.

While subsonic inviscid drag has little practical applications in itself, the solution can be used as a substitute for the numerical adjoint solution or to verify numerical adjoint solutions. For lifting solutions, on the other hand, the representation is only approximate but the agreement with the numerical adjoint solution is reasonable (especially for the wall-normal adjoint velocity) so that the benchmark solution can be used for a rough verification of the adjoint solution.

## Acknowledgements

The research described in this paper has been supported by INTA and the Ministry of Defence under the grant Termofluidodinámica (IGB99001). The computations reported in the paper were carried out with DLR's TAU Code, which is licensed to INTA through an R+D cooperation agreement, and with Stanford University's open-source code SU2.

## References

[1] Jameson A., "Aerodynamic Design via Control Theory," J. Sci. Comp., Vol. 3, No. 3, 1988, pp. 233-260. https://doi.org/10.1007/BF01061285

[2] Venditti D. and Darmofal D., "Grid Adaptation for Functional Outputs: Application to Two-Dimensional Inviscid Flows," Journal of Computational Physics, Vol. 176, 2002, pp. 40–69. https://doi.org/10.1006/jcph.2001.6967

[3] Luchini P. and Bottaro A., "Adjoint Equations in Stability Analysis," Ann. Rev. Fluid Mech., vol. 46, 2014, pp. 493-517. https://doi.org/10.1146/annurev-fluid-010313-141253

[4] Nadarajah S. and Jameson A., "A Comparison of the Continuous and Discrete Adjoint Approach to Automatic Aerodynamic Optimization," 38th AIAA Aerospace Sciences Meeting and Exhibit, AIAA paper 2000–0667, 2000. https://doi.org/10.2514/6.2000-667

[5] Lozano C. and Ponsin J., "Remarks on the Numerical Solution of the Adjoint Quasi-One-Dimensional Euler Equations," Int. J. Numer. Meth. Fluids, Vol. 69, No. 5, 2012, pp. 966-982. https://doi.org/10.1002/fld.2621

[6] Giles M., Duta M., Müller J.-D. and Pierce N., "Algorithm Developments for Discrete Adjoint Methods," AIAA Journal, Vol. 41, No. 2, 2003, pp. 198-205. https://doi.org/10.2514/2.1961

[7] Giles M. and Pierce N., "Analytic adjoint solutions for the quasi-one-dimensional Euler equations," J. Fluid Mechanics, Vol. 426, 2001, pp. 327-345. https://doi.org/10.1017/S0022112000002366

[8] Lozano C., "Singular and Discontinuous Solutions of the Adjoint Euler Equations," AIAA Journal, vol. 56, no. 11, 2018, pp. 4437-4452. https://doi.org/10.2514/1.J056523

[9] Fidkowski K. and Roe P., "An Entropy Adjoint Approach to Mesh Refinement," SIAM Journal on Scientific Computing, Vol. 32, No. 3, 2010, pp. 1261-1287. https://doi.org/10.1137/090759057

[10] Lozano C., "Entropy and Adjoint Methods," J Sci Comput, Vol. 81, 2019, pp. 2447–2483. https://doi.org/10.1007/s10915-019-01092-0.

[11] Anderson W.K. and Venkatakrishnan V., "Aerodynamic Design Optimization on Unstructured Grids with a Continuous Adjoint Formulation," Comput. Fluids, Vol. 28, 1999, pp. 443-480. https://doi.org/10.1016/S0045-7930(98)00041-3

[12] Castro C., Lozano C., Palacios F. and Zuazua E., "Systematic Continuous Adjoint Approach to Viscous Aerodynamic Design on Unstructured Grids," AIAA Journal, Vol. 45, No. 9, 2007, pp. 2125-2139. https://doi.org/10.2514/1.24859

[13] Kavvadias I., Papoutsis-Kiachagias E. and Giannakoglou K., "On the proper treatment of grid sensitivities in continuous adjoint methods for shape optimization," J. Comput. Phys., Vol. 301, 2015, pp. 1-18. https://doi.org/10.1016/j.jcp.2015.08.012

[14] Schwamborn D, Gerhold T. and Heinrich R., "The DLR TAU-Code: Recent Applications in Research and Industry," ECCOMAS CFD 2006 Conference, 2006.

[15] Giles M. and Pierce N., "Adjoint Equations in CFD: Duality, Boundary Conditions and Solution Behavior," 13th Computational Fluid Dynamics Conference, AIAA Paper 97–1850, June 1997. https://doi.org/10.2514/6.1997-1850

[16] Thomas J. and Salas M., "Far-field boundary conditions for transonic lifting solutions to the Euler equations," AIAA Journal, Vol. 24, No. 7, 1986, pp. 1074-1080. https://doi.org/10.2514/3.9394

[17] Fidkowski K., Ceze M. and Roe P., "Entropy-based Drag Error Estimation and Mesh Adaptation in Two Dimensions," Journal of Aircraft, Vol. 49, No. 5, 2012, pp. 1485-1496, 2012. https://doi.org/10.2514/1.C031795

[18] Economon T., Palacios F., Copeland S., Lukaczyk T. and Alonso J., "SU2: An Open-Source Suite for Multiphysics Simulation and Design," AIAA Journal, Vol. 54, No. 3, 2016, pp. 828-846. https://doi.org/10.2514/1.J053813